\documentclass[%
 aip,
 jcp,%
 amsmath,amssymb,
reprint,%
]{revtex4-2}

\usepackage{graphicx}
\usepackage{dcolumn}
\usepackage{bm}
\usepackage[usenames,dvipsnames]{color}

\usepackage{cancel}

\usepackage[dvipsnames]{xcolor}

\newcommand{\seeSI}{(see SI)}

\usepackage[normalem]{ulem}

\usepackage{tabularx, threeparttable}
\newcolumntype{L}[1]{>{\raggedright\arraybackslash}p{#1}} 
\newcolumntype{C}[1]{>{\centering\arraybackslash}p{#1}} 
\newcolumntype{R}[1]{>{\raggedleft\arraybackslash}p{#1}} 

\begin{document}

\setlength\parindent{0pt}


\title[]{Internal Trajectories and Observation Effects in Langevin Splitting Schemes}%

\author{Bettina G. Keller}
\email[]{bettina.keller@fu-berlin.de}
\affiliation{Department of Biology, Chemistry, Pharmacy, Freie Universit\"{a}t Berlin, Arnimallee 22, D-14195 Berlin, Germany}

\date{\today}

%
%


\begin{abstract}
Langevin integrators based on operator splitting are widely used in molecular dynamics.
This work examines Langevin splitting schemes from the perspective of their internal trajectories and observation points, complementing existing generator-based analyses.
By exploiting merging, splitting, and cyclic permutation of elementary update operators, formally distinct schemes can be grouped according to identical or closely related trajectories.
Accuracy differences arising from momentum updates and observation points are quantified for configurational sampling, free-energy estimates, and transition rates.
While modern Langevin integrators are remarkably stable under standard simulation conditions, subtle but systematic biases emerge at large friction coefficients and time steps. 
These results clarify when accuracy differences between splitting schemes matter in practice and provide an intuitive framework for understanding observation effects.
\end{abstract}

\maketitle


\section{Introduction}
Molecular dynamics (MD) is a powerful tool for simulating atomistic systems, but realistic simulations typically require temperature control.
Temperature is controlled in the MD integrator by modifying the momenta. 
Conceptually the momenta are coupled to a heat bath, such that the resulting momentum statistics are consistent with a Maxwell–Boltzmann distribution at the target temperature \cite{frenkel2023understanding, leimkuhler2015molecular}.
Thermostat algorithms exist in both deterministic and stochastic variants \cite{hunenberger2005thermostat}. 
Thermostat algorithms exist in deterministic and stochastic variants.

Among the latter, Langevin thermostats occupy a special role because they can be formulated at the level of a stochastic differential equation, with a well-defined generator whose discretization directly leads to practical integration schemes.
This aspect is particularly relevant in transition path sampling, developed by Dellago \textsl{et.~al.}\cite{dellago1998efficient, dellago1998transition, bolhuis2002transition}, and in trajectory reweighting techniques \cite{keller2024dynamical}, where molecular dynamics trajectories themselves become the object of statistical analysis.
In such approaches, the thermostat determines the underlying dynamical model and its associated path probability measure, and trajectory-level observables can be highly sensitive to this choice \cite{van1998largest, dellago1997lyapunov, ceriotti2011accelerating}.
For Langevin dynamics, the path measure is well defined both in continuous time and for the corresponding discrete-time dynamics generated by numerical integrators, enabling the explicit construction of path probabilities and reweighting factors \cite{Donati:2018, Kieninger:2021, Sivak:2014}.
Langevin thermostat schemes have been derived from a wide range of starting points \cite{Brunger:1984, van1988leap, pastor1988analysis, izaguirre2001langevin, bou2009stochastic, Bussi:2007, melchionna2007design, Izaguirre:2010, Goga:2012, Leimkuhler:2012, gronbech2020complete}.
A particularly useful class of methods is based on operator splitting\cite{Sexton:1992, Tuckerman:1992, bou2009stochastic, Bussi:2007, Leimkuhler:2012}, in which the generator of the dynamics is split into additive terms and the propagator is approximated using the Baker--Campbell--Hausdorff expansion \cite{suzuki1976generalized}.
%
%
Langevin splitting schemes use force information from a single time level per integration step, although leap-frog-like non-splitting variants have also been proposed \cite{jensen2019accurate, gronbech2020complete, finkelstein2021bringing}.
Despite different starting points of the derivation, several Langevin schemes have been shown to be equivalent - either exactly or in an appropriate sense - showing that apparently distinct algorithms can generate identical or closely related dynamics \cite{Leimkuhler:2013, Sivak:2014, finkelstein2021bringing, kieninger2022gromacs}.
Furthermore, Langevin integrators can be categorized by the placement of the thermostat step in the integration cycle \cite{Zhang:2019}, and recent developments have been summarized in a dedicated review \cite{agarwal2025advanced}.
In the following, I focus on Langevin splitting schemes, as they have been analyzed in considerable detail \cite{BouRabee:2010, Leimkuhler:2012, Leimkuhler:2013, Sivak:2013, Sivak:2014, Fass:2018, lelievre2016partial} and are particularly well suited for the explicit construction and analysis of path actions.
A rigorous analysis of Langevin schemes - covering properties such as ergodicity, time reversibility, and convergence - needs to be formulated on the level of probability distributions and their time-evolution. 
This leads to a framework based on Liouville and Fokker-Planck equations that offers deep structural insight into deterministic and stochastic dynamics.
At the same time, most MD practitioners approach integration schemes from the perspective of the equations of motion for individual particles.
This viewpoint is reinforced by standard MD curricula, where integration schemes are commonly introduced via Taylor expansions of the equations of motion. 
More importantly, in the development, implementation, and maintenance of MD software, one works directly with the update rules themselves. 
This requires a strong understanding of how these update rules translate into the trajectories.
While distribution-level descriptions of Langevin schemes are essential for a full grasp of these algorithms, many practically relevant aspects of convergence can already be understood directly at the level of the integrator equations themselves.
In practical simulations, positions and momenta are rarely written to disk at every integration step; instead, hundreds or thousands of steps are typically performed between output frames.
From this perspective, the internal trajectories generated between write-out points play a central role, yet are seldom examined explicitly.
In this work, I therefore focus on analyzing and comparing these internal trajectories, and show that this viewpoint provides valuable insight into the behavior and accuracy of different Langevin splitting schemes.

\section{Theory}
\label{sec:Theory}

\begin{table*}[t]
    \centering
    \begin{tabular}{ll|l|l|l}
    \hline 
    \multicolumn{4}{c|}{ABO schemes} & AP schemes\\[0.6ex]
    \hline    
    \rule{0pt}{2.6ex}%
    symmetric &&non-symmetric  &first-order    &symmetric \\[0.6ex]
    \hline
    &&&\\[-0.2cm]
    $\;\; \mathcal{U}_{BAOAB} = \mathcal{B'}\mathcal{A'}\mathcal{O}\mathcal{A'}\mathcal{B'} \;\;$&
    $\;\; \mathcal{U}_{OABAO} = \mathcal{O'}\mathcal{A'}\mathcal{B}\mathcal{A'}\mathcal{O'} \;\;$&
    $\;\; \mathcal{U}_{BAOA} = \mathcal{A'}\mathcal{O}\mathcal{A'}\mathcal{B} \;\;$&
    $\;\; \mathcal{U}_{ABO} = \mathcal{O}\mathcal{B}\mathcal{A} \;\;$&
    $\;\; \mathcal{U}_{APA} = \mathcal{A'}\mathcal{P}\mathcal{A'} \;\;$\\[0.6ex]
    $\;\; \mathcal{U}_{ABOBA} = \mathcal{A'}\mathcal{B'}\mathcal{O}\mathcal{B'}\mathcal{A'} \;\;$&
    $\;\; \mathcal{U}_{OBABO} = \mathcal{O'}\mathcal{B'}\mathcal{A}\mathcal{B'}\mathcal{O'} \;\;$ &
    $\;\; \mathcal{U}_{ABOB} = \mathcal{B'}\mathcal{O}\mathcal{B'}\mathcal{A} \;\;$&
    $\;\; \mathcal{U}_{AOB} = \mathcal{B}\mathcal{O}\mathcal{A} \;\;$&
    $\;\; \mathcal{U}_{PAP} = \mathcal{P'}\mathcal{A}\mathcal{P'} \;\;$\\[0.6ex]
    $\;\; \mathcal{U}_{BOAOB} = \mathcal{B'}\mathcal{O'}\mathcal{A}\mathcal{O'}\mathcal{B'} \;\;$&
    $\;\; \mathcal{U}_{AOBOA} = \mathcal{A'}\mathcal{O'}\mathcal{B}\mathcal{O'}\mathcal{A'} \;\;$&
    $\;\; \mathcal{U}_{BOAO} = \mathcal{O'}\mathcal{A}\mathcal{O'}\mathcal{B} \;$&&\\[0.6ex]
    \hline
    \end{tabular}
    \caption{Langevin splitting schemes}
    \label{tab:LangevinSchemes}
\end{table*}

\subsection{From Langevin dynamics to update operators}
Consider a particle with mass $m$ moving in one-dimensional position space $q \in \mathbb{R}$ according to underdamped Langevin dynamics
\begin{align}
     \binom{\dot q}{\dot p}&=
       \underbrace{\binom{p/m}{0}}_{A} +
       \underbrace{\binom{0}{- \nabla_q V(q)}}_{B} +
       \underbrace{\binom{0}{ -\xi p +  \sigma\, \eta(t)} }_{O} \, . 
 \label{eq:underdamped_Langevin_01}         
\end{align}
In eq.~\ref{eq:underdamped_Langevin_01}, 
$V(q)$ is the potential energy function,
$\nabla_q = \partial/\partial q$ denotes the gradient with respect to the position coordinate,
$\xi$  is a collision or friction rate (in units of s$^{-1}$),
$\sigma =\sqrt{2\xi k_B T m}$, where
$T$ is the temperature and $k_B$ is the  Boltzmann constant. 
The stochastic process $\eta \in \mathbb{R}$ denotes Gaussian white noise with zero mean and unit variance,
$\langle \eta(t) \rangle =0$ and  $\langle \eta(t)\eta(t') \rangle = \delta(t-t')$,
where $\delta(t-t')$ is the Dirac delta-function.
We use the dot-notation for time derivatives: $\dot q = \partial q / \partial t$.
$(q(t), p(t)) \subset \mathbb{R}^2$ denotes the state of the system at time $t$, which consists of positions $q(t)$ and momenta $p(t) = m\dot{q}(t)$.
Eq.~\ref{eq:underdamped_Langevin_01} splits the vector field into three terms: $A$, $B$ and $O$.
$A$ affects the position $q$, while $B$ and $O$ affect the momentum $p$, where $O$ contains the stochastic process $\eta(t)$.
The deterministic parts $A$ and $B$ represent ordinary differential equations with analytical solutions.
The stochastic part $O$ represents an Ornstein-Uhlenbeck process, 
$\dot p = -\xi p + \sigma\eta(t)$, which is likewise exactly solvable \cite{Leimkuhler:2016}. 
From these analytical solutions one can construct the following elementary update operators \cite{Leimkuhler:2016}
\begin{subequations}
\begin{align}
     \mathcal{A}
     \begin{pmatrix} q_{k}\\p_{k}\end{pmatrix}
     &=  \begin{pmatrix} q_k+a p_k \\p_k\end{pmatrix}\label{eq:A}\\
     \mathcal{B}\begin{pmatrix} q_{k}\\p_{k}\end{pmatrix}
     &= \begin{pmatrix}q_k \\ p_k+b(q_k)\end{pmatrix}\label{eq:B}\\
     \mathcal{O}\begin{pmatrix} q_{k}\\ p_{k}\end{pmatrix}
     &= \begin{pmatrix} q_k \\ d \, p_k + f\, \eta_k\end{pmatrix}\label{eq:O}\, ,
\end{align}
\end{subequations}
with parameters
\begin{subequations}
\begin{align}
    a       &= \Delta t \frac{1}{m} \label{eq:a} \\
    b(q_k)  &= -\Delta t\nabla V(q_k) \label{eq:b} \\
    d       &= e^{-\xi\Delta t}  \label{eq:d} \\
    f       &= \sqrt{k_BTm(1-e^{-2\xi\Delta t})}  \label{eq:f}  \, ,
\end{align}
\end{subequations}
where $\Delta t$ is the time step and $\eta_k$ is a normally distributed random number with zero mean $\mu$ and unit variance $\sigma^2$, $\eta_k \sim \mathcal{N}(\mu,\sigma^2) = \mathcal{N}(0,1)$.
The parameter names $a$ and $b$ correspond to the operators $\mathcal{A}$ and $\mathcal{B}$, respectively, while $d$ and $f$ denote dissipation and fluctuation in the Ornstein–Uhlenbeck step.  
In the standard terminology, the $\mathcal{A}$-update is referred to as the drift step, the $\mathcal{B}$-update as the force (or kick) step, and the $\mathcal{O}$-update as the thermostatting step.
In eqs.~\ref{eq:A} to \ref{eq:O}, the initial state $(q(t), p(t)) = (q_k, p_k)$ is propagated over a time interval $\Delta t$. 
The elementary update operators can be chained in different sequences yielding a family of MD integrators called Langevin splitting schemes
\cite{BouRabee:2010, Bussi:2007, Leimkuhler:2012, Leimkuhler:2013, Fass:2018, Sivak:2013, Zhang:2019}.
If an elementary update operator is carried out for only half a time step $\frac{\Delta t}{2}$, it is denoted with a prime, e.g.
\begin{align}
     \mathcal{A'}\left(\begin{array}{c}q_{k}\\ p_{k}\end{array}\right) &= \left(\begin{array}{c}q_k+\frac{\Delta t}{2}\frac{1}{m}p_k \\ p_k\end{array}\right) 
     = \left(\begin{array}{c} q_k+a' p_k \\p_k\end{array}\right)\, .
\end{align}
Correspondingly, $a'$, $b'$, $d'$ and $f'$ are obtained by replacing $\Delta t$ with $\frac{\Delta t}{2}$ in eqs.~\ref{eq:a}-\ref{eq:f}.

%
\subsection{Langevin integrators based on ABO-update operators}

The derivation and analysis of specific update operator sequences is carried out most rigorously in the generator formulation.
In this framework, the stochastic differential equation (eq.~\ref{eq:underdamped_Langevin_01}) is associated with its (backward) Kolmogorov generator
\(\mathcal{L} = \mathcal{L}_A + \mathcal{L}_B + \mathcal{L}_O\), where \(\mathcal{L}_A\), \(\mathcal{L}_B\), and \(\mathcal{L}_O\) are the generators corresponding to the vector fields \(A\), \(B\), and \(O\), respectively.
This decomposition is exact, but the corresponding time-evolution operator $\exp(\Delta t\,\mathcal{L})$ cannot, in general, be factorized exactly into a product of the individual propagators. 
Instead, Lie-Trotter and Trotter-Strang splittings are employed to construct systematic approximations of $\mathcal{L}$.
These operator splittings translate directly into concrete sequences of update operators acting on individual trajectories \cite{bou2009stochastic, bou2010pathwise, leimkuhler2016computation, lelievre2016partial, Sivak:2014}.  
In the present work, I deliberately remain at the level of the explicit update operators and the operations that are carried during an MD simulation, and examine how much insight can be gained directly from this representation.
The general equation for a single MD integration step is
\begin{align}
    \mathcal{U}_{\mathrm{scheme}} \begin{pmatrix} q_{k}\\p_{k}\end{pmatrix}  &= \begin{pmatrix} q_{k+1}\\p_{k+1}\end{pmatrix} 
\end{align}
where $\mathcal{U}_{\mathrm{scheme}}$ denotes the update operator for the full MD step and the subscript "scheme" specifies the splitting scheme on which the operator is based.
$k$ denotes the index of the integration step, such that time $t = k \Delta t$.
\paragraph{First-order ABO schemes}
First-order integration schemes correspond to sequentially applying the three update operators $A$, $B$, and $O$ over a full time step $\Delta t$.
For example, the ABO scheme corresponds to the update operator
\begin{align}
    \mathcal{U}_{\mathrm{ABO}} \binom{q_k}{p_k} &= \mathcal{O}\mathcal{B}\mathcal{A} \binom{q_k}{p_k} = \binom{q_{k+1}}{p_{k+1}} \, 
\end{align}
where the elementary update operators are applied right-to-left, while scheme name is read left-to-right.
As a result, for non-symmetric schemes, the order of operators is reversed relative to the letters in the scheme name.
Such schemes correspond to Lie–Trotter splittings and are first-order accurate in the weak sense.

In principle, there are $3! = 6$ distinct orderings of the update operators.
However, cyclic permutations of a given sequence lead to equivalent schemes, since they correspond only to a shift of the starting point (or ``reading frame'') along a periodically repeated operator sequence:
\begin{align}
    (ABO)(ABO)(ABO)\dots    &= A(BOA)(BOA)BO\dots \cr
                            &= AB(OAB)(BOA)O\dots
\label{eq:ABO_cyclicPermutation}      
\end{align}
which shows that ABO, BOA, and OAB differ only by a cyclic shift of the operator sequence.
As a result, the six possible orderings of $\mathcal{A}$, $\mathcal{B}$, and $\mathcal{O}$ reduce to two distinct equivalence classes of first-order integrators (see Fig.~\ref{fig:LangevinSplitting_01}d and Table~\ref{tab:LangevinSchemes}).
%
%
%


\paragraph{Non-symmetric ABO schemes.}
Non-symmetric second-order schemes arise from Strang-type compositions of the form $XY'ZY'$ and $ZY'XY'$, with $X,Y,Z \in \{\mathcal{A},\mathcal{B},\mathcal{O}\}$, $X\neq Y\neq Z$, and the prime denoting a half-step.  
With two sequence types and $3!=6$ possible operator assignments, this yields 12 schemes.  
Since each four-operator sequence has four cyclic permutations (e.g. $BAOA \sim AOAB \sim OABA \sim ABAO$), these reduce to three distinct equivalence classes (Tab.~\ref{tab:LangevinSchemes}).
%
%


\paragraph{Symmetric ABO schemes.}
Symmetric second-order schemes arise from Strang-type composition of the form $X'Y'ZY'X'$, with $X,Y,Z \in \{\mathcal{A},\mathcal{B},\mathcal{O}\}$, $X\neq Y\neq Z$, and the prime denoting a half-step.  
Formally, there are $3! = 6$ such compositions (Tab.~\ref{tab:LangevinSchemes}).
%
%
The OBABO integration scheme can be viewed as a limiting case of the Bussi–Parrinello thermostat, in which the thermostat is applied independently to each degree of freedom \cite{Bussi:2007}.

\subsection{Langevin integrators based on AP-update operators}
An alternative way to decompose eq.~\ref{eq:underdamped_Langevin_01} is to split it into a vector field $A$, which governs the time-evolution of the positions, and a vector field $P$, which governs the time-evolution of the momenta\cite{melchionna2007design}
\begin{align}
     \binom{\dot q}{\dot p} &=
     \underbrace{\binom{p/m}{0}}_{A} +
     \underbrace{\binom{0}{- \nabla_q V(q)  -\xi p + \sigma \, \eta(t)} }_{P} \, .
\label{eq:underdamped_Langevin_02}         
\end{align}
The update operator for $A$ is given by eq.~\ref{eq:A}. 
The vector field $P$ combines the $B$- and $O$-vector field into a single stochastic differential equation for the momenta: 
\begin{align}
     \dot p &= - \nabla_q V(q) -\xi p + \sigma\, \eta(t) \, .
\label{eq:OU_with_drift}     
\end{align}
Eq.~\ref{eq:OU_with_drift} describes an Ornstein–Uhlenbeck process with an additional deterministic drift term arising from the conservative force.
During a $\mathcal{P}$-update the position is held fixed, so that $-\nabla_q V(q)$ is constant over the interval $[t,t+\Delta t]$.
Under this condition, eq.~\ref{eq:OU_with_drift} admits the following exact update\cite{Leimkuhler:2016}:
\begin{align}
    \mathcal{P}\left(\begin{array}{c} q_{k}\\p_{k}\end{array}\right) 
    &=
    \begin{pmatrix}
        q_k \\
        \frac{1 - d}{\xi\Delta t} \, b(q_k)+ d p_k  + f \, \eta_k
    \end{pmatrix} 
    \, ,
\label{eq:P}     
\end{align}
where $\eta_k \sim \mathcal{N}(0,1)$ is a normally distributed random number with zero mean and unit variance, and parameters $b(q_k)$, $d$ and $f$ are defined in eqs.~\ref{eq:b}, \ref{eq:d} and \ref{eq:f}.
As before, a prime denotes that the udpate operator is carried out for half a time step.
The AP-splitting of the vector field gives rise to two symmetric schemes (Tab.~\ref{tab:LangevinSchemes}).
%
The scheme $APA$ is the Stochastic Position Verlet integration method, whereas $PAP$ is the Stochastic Velocity Verlet integration method\cite{melchionna2007design}.
%

\subsection{Cyclic permutation}
MD trajectories are typically recorded at sparse intervals, for example every 100 or 1000 integration steps.
It is therefore useful to distinguish the \emph{internal trajectory}, i.e. the intermediate steps that are generated by the integrator, from \emph{recorded trajectory} which is written to disk. 
Cyclic permutations of a given operator sequence do not change the internal trajectory. 
Instead, they have two distinct effects:
($i$) they shift the entry point into the periodically repeated integration cycle by a finite number of update operations, and
($ii$) they shift the point within the cycle at which positions and momenta are recorded.
\begin{figure*}[t]
    \includegraphics[width=\textwidth]{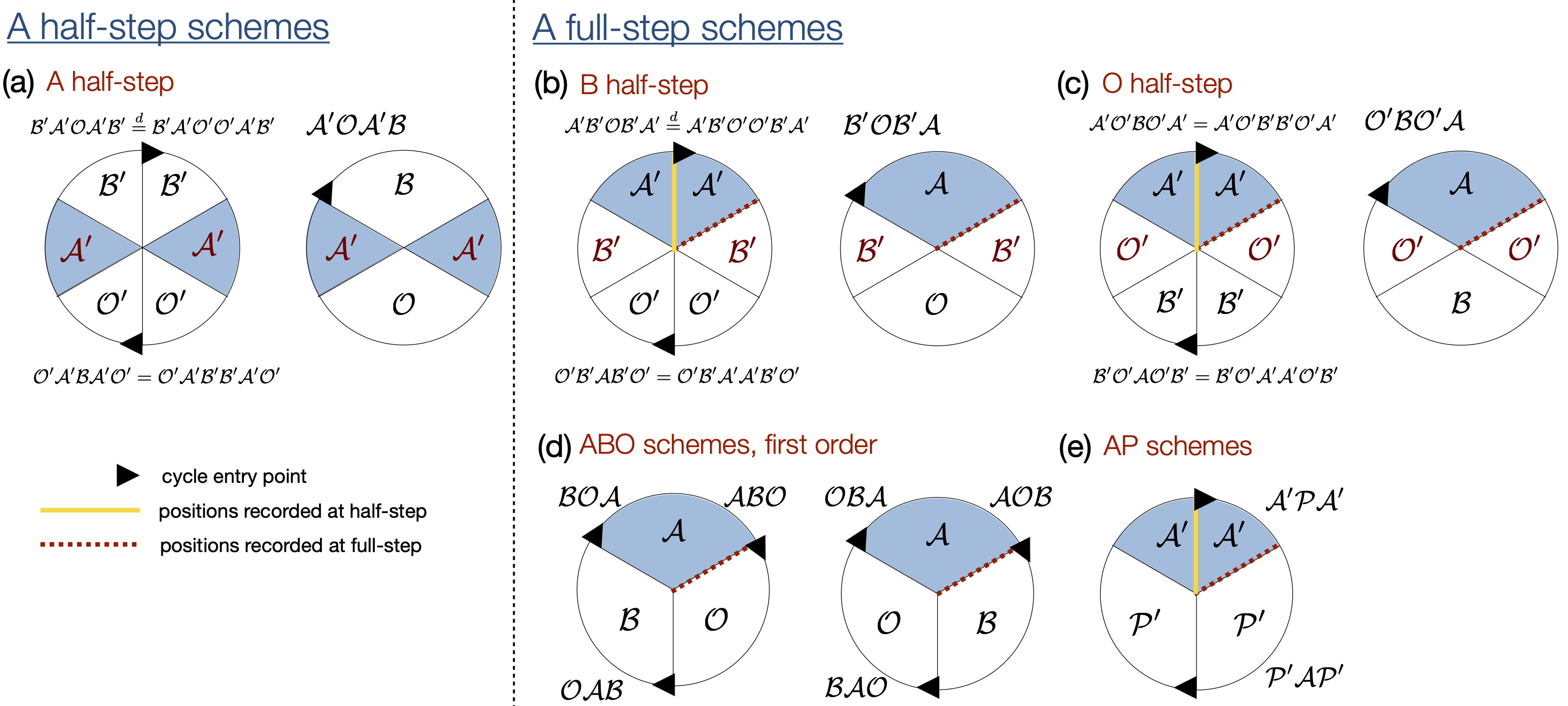}
\caption{Visualization and classification of Langevin splitting operators. Elementary update operators are applied right to left.
} 
\label{fig:LangevinSplitting_01}
\end{figure*}

Eq.~\ref{eq:ABO_cyclicPermutation} illustrates this: the operator sequences
$(\mathcal{O}\mathcal{B}\mathcal{A})^n$ and $\mathcal{O}\mathcal{B}(\mathcal{A}\mathcal{O}\mathcal{B})^{n-1}\mathcal{A}$ are identical.
The first effect can be interpreted as a finite shift of the initial condition \cite{Leimkuhler:2013, leimkuhler2015molecular}.
Specifically, if $(q_0,p_0)$ is used as the initial condition for the ABO scheme, then choosing $\mathcal A(q_0,p_0)^{\top}$ as the initial condition for the BOA scheme yields the same internal trajectory, provided that the same sequence of random numbers is employed \cite{kieninger2022gromacs}.
For stochastic integration schemes, such finite shifts of the initial conditions do not affect convergence properties or long-time statistics \cite{leimkuhler2016computation}.
Consequently, this first effect does not alter the accuracy of the scheme.
The second effect, which concerns the observation point within the integration cycle, can however significantly influence the accuracy of the measured quantities and will be analyzed in detail in Sec.~\ref{sec:Position_update}.

%

\section{Results}
Figure~\ref{fig:LangevinSplitting_01} offers a visualization for the 13 Langevin integration schemes in Tab.~\ref{tab:LangevinSchemes}. 
The schemes are grouped according to their update operator sequences.

\subsection{Merging and splitting elementary update operators}
When analyzing Langevin schemes at the level of their internal trajectories, a useful observation is that applying an elementary update operator ($\mathcal{A}$, $\mathcal{B}$, $\mathcal{O}$, or $\mathcal{P}$) for a full time step gives the same result as applying it twice with half the time step.
For the position update $\mathcal{A}$ and the conservative force update $\mathcal{B}$, this equivalence holds exactly 
\seeSI
\begin{align}
    \mathcal{A'}\mathcal{A}' =   \mathcal{A}, \quad
    \mathcal{B'}\mathcal{B}' =   \mathcal{B}  \, .
\label{eq:AA_and_BB}     
\end{align}
For the thermostatting step $\mathcal{O}$ this is not immediately obvious, because two half-step updates require two random numbers, whereas a full-step update only uses a single random number.
For two half-steps one obtains
\begin{align}
        \mathcal{O'}\mathcal{O}' \begin{pmatrix} q_k \\ p_k \end{pmatrix}
    &=   \begin{pmatrix} q_k\\ d \, p_k + f'\left(d'\, \eta_k^{(1)} +  \eta_k^{(2)}\right)\end{pmatrix}
\end{align}
where $\eta_k^{(1)},\eta_k^{(2)}\sim\mathcal N(0,1)$ are independent normally distributed random variables.
This update is statistically equivalent to a single full $\mathcal O$ step
(Eq.~\ref{eq:O}) provided that the full-step noise $\eta_k$ satisfies $f \eta_k = f'\!\left(d'\, \eta_k^{(1)} +  \eta_k^{(2)} \right)$. 
Accordingly, $\eta_k$ can be written as the weighted sum
\begin{align}
\eta_k
=
\frac{f'}{f} d'\,\eta_k^{(1)}
+
\frac{f'}{f}\,\eta_k^{(2)} .
\label{eq:O_eta_k}
\end{align}
Since $\eta_k^{(1)}$ and $\eta_k^{(2)}$ are independent standard normal
random variables, $\eta_k$ is also normally distributed with
\begin{align}
\mu = 0,
\qquad
\sigma^2
=
\left(\frac{f'}{f} d'\right)^2
+
\left(\frac{f'}{f}\right)^2
= 1 .
\end{align}
Thus, $\eta_k\sim\mathcal N(0,1)$.
This construction shows that the two independent Gaussian random variables used in the half-step updates can be mapped onto a single Gaussian random variable for the full step, and vice versa. While two half-step updates are not identical to a full-step update at the level of individual trajectories, they generate the same momentum distribution. Consequently, on the level of
distributions,
\begin{align}
\mathcal O'\mathcal O' &\stackrel{d}{=} \mathcal O ,
\label{eq:OO}
\end{align}
where $\stackrel{d}{=}$ denotes equality in distribution.
(This argument can equivalently be made in terms of the time-evolution operator \cite{finkelstein2021bringing}.)

\begin{figure}
    \centering
    \includegraphics[width=\columnwidth]{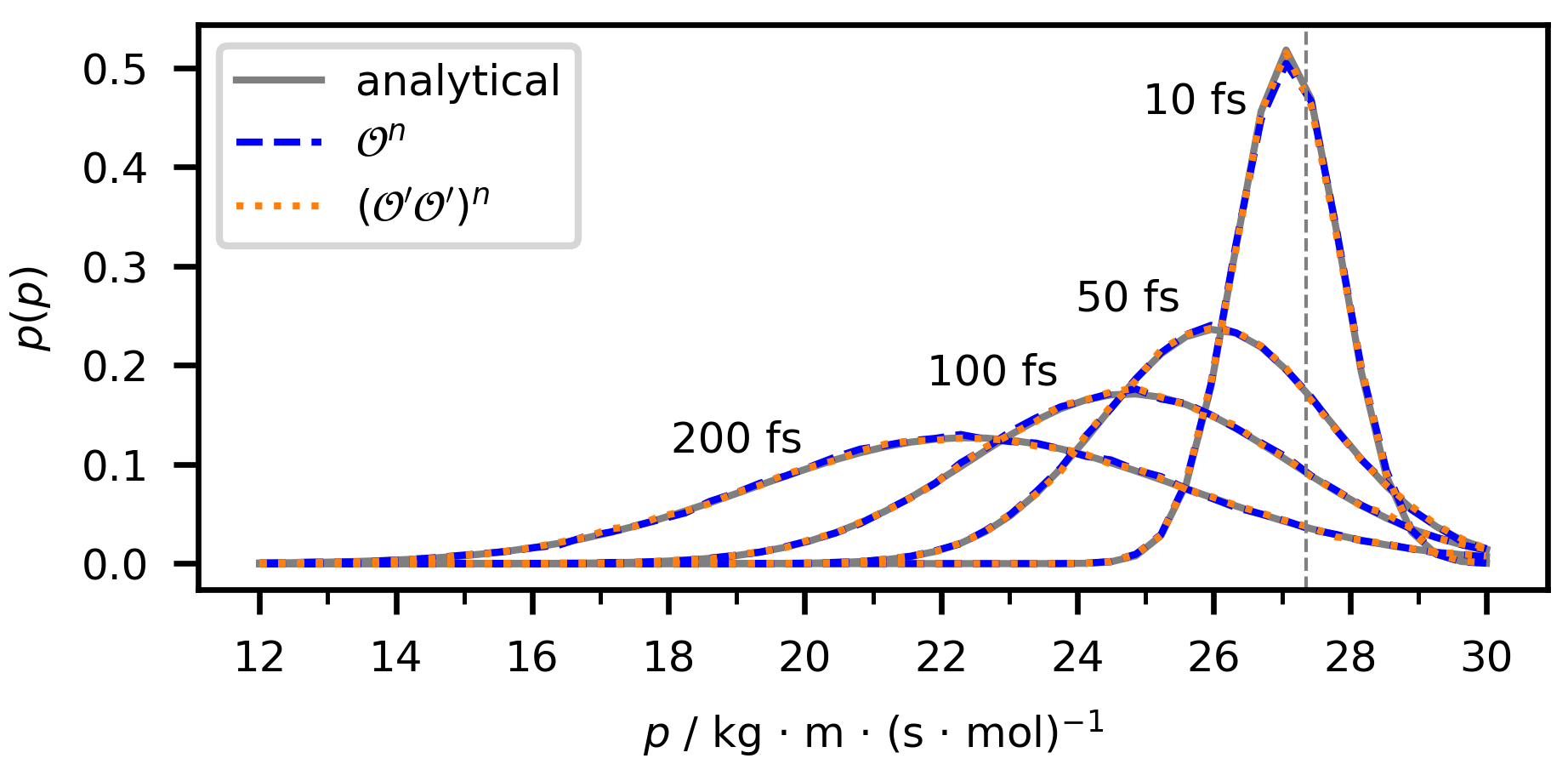}
    \caption{Momentum distributions of a mass $m=12.0\, \mathrm{g}/\mathrm{mol}$ 
    at $T=300\, \mathrm{K}$ and $\xi = 1\, \mathrm{ps}^{-1}$
    obtained by propagating with full O-steps and with half O-steps compared to the analytical momentum distribution
    Vertical dashed line indicates the initial condition $p_0$.
    }
    \label{fig:O_step}
\end{figure}
Fig.~\ref{fig:O_step} shows that the relation in eq.~\ref{eq:OO} holds up in a numerical experiment.
The test particle is a carbon atom with mass $m = 12 \mathrm{g}/\mathrm{mol}$ at 300 K.
In the simulation, a swarm of trajectories is initialized with the same initial momentum momentum $p_0$.
The momentum distribution obtained by propagating the trajectories with full-step updates $\mathcal{O}^n$ (dashed blue lines) match those obtained by propagating with half-step updates $(\mathcal{O}'\mathcal{O}')^n$ (dotted orange lines) exactly,  where $n$ is the number of iterations.
Moreover, both distributions coincide with the analytical distribution (gray line).
By an analogous argument \seeSI, two half-step $\mathcal{P}$ updates are equivalent in distribution to a single full-step update.
\begin{align}
\mathcal P'\mathcal P' &\stackrel{d}{=} \mathcal P .
\label{eq:PP}
\end{align}
In brief, merging and splitting of elementary update steps does not alter the statistical properties of the scheme.
\subsection{Symmetric and non-symmetric ABO schemes}
A direct consequence of the previous section is that the six symmetric ABO schemes can be grouped into three pairs, where each pair executes the same sequence of integration sub-steps and therefore generates the same internal trajectory.
Fig.~\ref{fig:LangevinSplitting_01}a illustrates this for the pair BAOAB and OABAO.
Since splitting elementary update operators does not alter the statistical properties of the operator sequence, the central full update may be replaced by two consecutive half updates:
$\mathcal{B}'\mathcal{A}'\mathcal{O}\mathcal{A}'\mathcal{B}' \stackrel{d}{=}
\mathcal{B}'\mathcal{A}'\mathcal{O}'\mathcal{O}'\mathcal{A}'\mathcal{B}'$ and similarly 
$\mathcal{O}'\mathcal{A}'\mathcal{B}\mathcal{A}'\mathcal{O}' \stackrel{d}{=}
\mathcal{O}'\mathcal{A}'\mathcal{B}'\mathcal{B}'\mathcal{A}'\mathcal{O}'$.
In this form, it becomes evident that BAOAB and OABAO are related by a cyclic permutation of the operator sequence and therefore differ only by the choice of entry point into the periodically repeated integration cycle.
In Fig.~\ref{fig:LangevinSplitting_01}.a, BAOAB enters the cycle at the top, whereas OABAO enters it at the bottom. 
Apart from this shift in the starting point, both schemes apply an identical sequence of update operations and thus generate the same internal trajectory. 
Analogously, ABOBA is a cyclic permutation of OBABO (Fig.~\ref{fig:LangevinSplitting_01}.b), and BOAOB is a cyclic permutation of AOBOA (Fig.~\ref{fig:LangevinSplitting_01}.c).
Each pair of symmetric ABO schemes can be related to a non-symmetric counter part by merging the two consecutive half-step updates at the cycle boundary. 
For example, in the BAOAB scheme (Fig.~\ref{fig:LangevinSplitting_01}.a), merging the final $\mathcal{B}'$ update of the $(k-1)$st iteration with the initial $\mathcal{B}'$ update of the $k$th iteration yields the BAOA scheme: $\dots \mathcal{B}'\;\; \mathcal{B}'\,\mathcal{A}'\,\mathcal{O}\,\mathcal{A}'\,\mathcal{B}'\;\; \mathcal{B}'\dots\,= \dots\mathcal{B}\,\mathcal{A}'\,\mathcal{O}\,\mathcal{A}' \;\;\mathcal{B}\dots$.
By the same argument, the schemes ABOBA and OBABO are equivalent to the the ABOB scheme, and BOAOB and AOBOA are equivalent to the BOAO scheme (Fig.~\ref{fig:LangevinSplitting_01}b and c).
(These equivalences can also be derived at the level of time-evolution operators \cite{Leimkuhler:2013,leimkuhler2015molecular}.)
This observation establishes a natural and systematic classification of second-order Langevin integrators according to which update operator retains a half-step character upon iteration:
$\mathcal{A}$-half-step schemes (Fig.~\ref{fig:LangevinSplitting_01}.a),  
$\mathcal{B}$-half-step schemes (Fig.~\ref{fig:LangevinSplitting_01}.b), and  
$\mathcal{O}$-half-step schemes (Fig.~\ref{fig:LangevinSplitting_01}.c).
\subsection{A-full-step schemes}
The schemes in Fig.~\ref{fig:LangevinSplitting_01}b and c differ from the $\mathcal{A}$-half-step schemes in that they effectively iterate full position updates together with full momentum updates.  
That is, their update sequence can be written in the generic form $(\mathcal{A}\mathcal{P}_{\mathrm{gen}})^n$, where $\mathcal{P}_{\mathrm{gen}}$ denotes a generic momentum update operator.
In this sense, they are closely related to the first-order schemes shown in Fig.~\ref{fig:LangevinSplitting_01}d, which likewise iterate full-step position and momentum updates.  
Similarly, the symmetric schemes APA and PAP are cyclic permutations of each other and share the same update structure $(\mathcal{A}\mathcal{P})^n$.
All schemes in Fig.~\ref{fig:LangevinSplitting_01}b--d are therefore structurally equivalent in that they generate trajectories by alternating full updates in position and momentum space.  
They differ only in the construction of the momentum update.

\subsection{Momentum update in a linear potential}
In Fig.~\ref{fig:LangevinSplitting_01}.b-d, the momentum update is constructed by combining $\mathcal{B}$- and $\mathcal{O}$-updates.
The resulting full-step momentum updates are \seeSI:
\begin{subequations}
\begin{align}
    \mathcal{B}' \mathcal{O}\mathcal{B}'\begin{pmatrix} q_k \\p_k\end{pmatrix} 
    &= \begin{pmatrix} q_k \\ \frac{d+1}{2} b(q_k) + dp_k + f \eta_k \end{pmatrix} \label{eq:BOB}\\
    \mathcal{O}' \mathcal{B}\mathcal{O}' \begin{pmatrix} q_k \\p_k\end{pmatrix}
    &= \begin{pmatrix} q_k \\ d'b(q_k) +  dp_k +f \eta_k \end{pmatrix} \label{eq:OBO}\\
    \mathcal{B} \mathcal{O} \begin{pmatrix} q_k \\p_k\end{pmatrix} 
    &=\begin{pmatrix} q_k \\  b(q_k) + dp_k  + f\eta_k\end{pmatrix} \label{eq:OB}\\
    \mathcal{O} \mathcal{B} \begin{pmatrix} q_k \\p_k\end{pmatrix}  
    &= \begin{pmatrix} q_k \\ db(q_k) + dp_k + f\eta_k\end{pmatrix} \label{eq:BO}
\end{align}
\end{subequations}
These composite momentum operators all define a full-step momentum update of the common form
\begin{align}
    p_{k+1} &= c\cdot b(q_k) + dp_k + f\eta_k \, 
\label{eq:generalMomentumUpdate}    
\end{align}
where the $\mathcal{O}$-step contributes the term $dp_k + f\eta_k$, and the prefactor $c$ scales the relative strength of the force kick ($\mathcal B$ step). 
This prefactor differs across the schemes.
To assess the accuracy of the four momentum update schemes, they can be compared to the exact update of the Ornstein-Uhlenbeck process with constant force (eq.~\ref{eq:OU_with_drift}), which is provided by the $\mathcal{P}$-update operator (eq.~\ref{eq:P}).
In the $\mathcal{P}$ update, the thermostatting step also contributes the term $dp_k + f\eta_k$, and the force kick enters with the prefactor $(1-d)/(\xi \Delta t)$.

\begin{figure}
    \centering
    \includegraphics[width=\columnwidth]{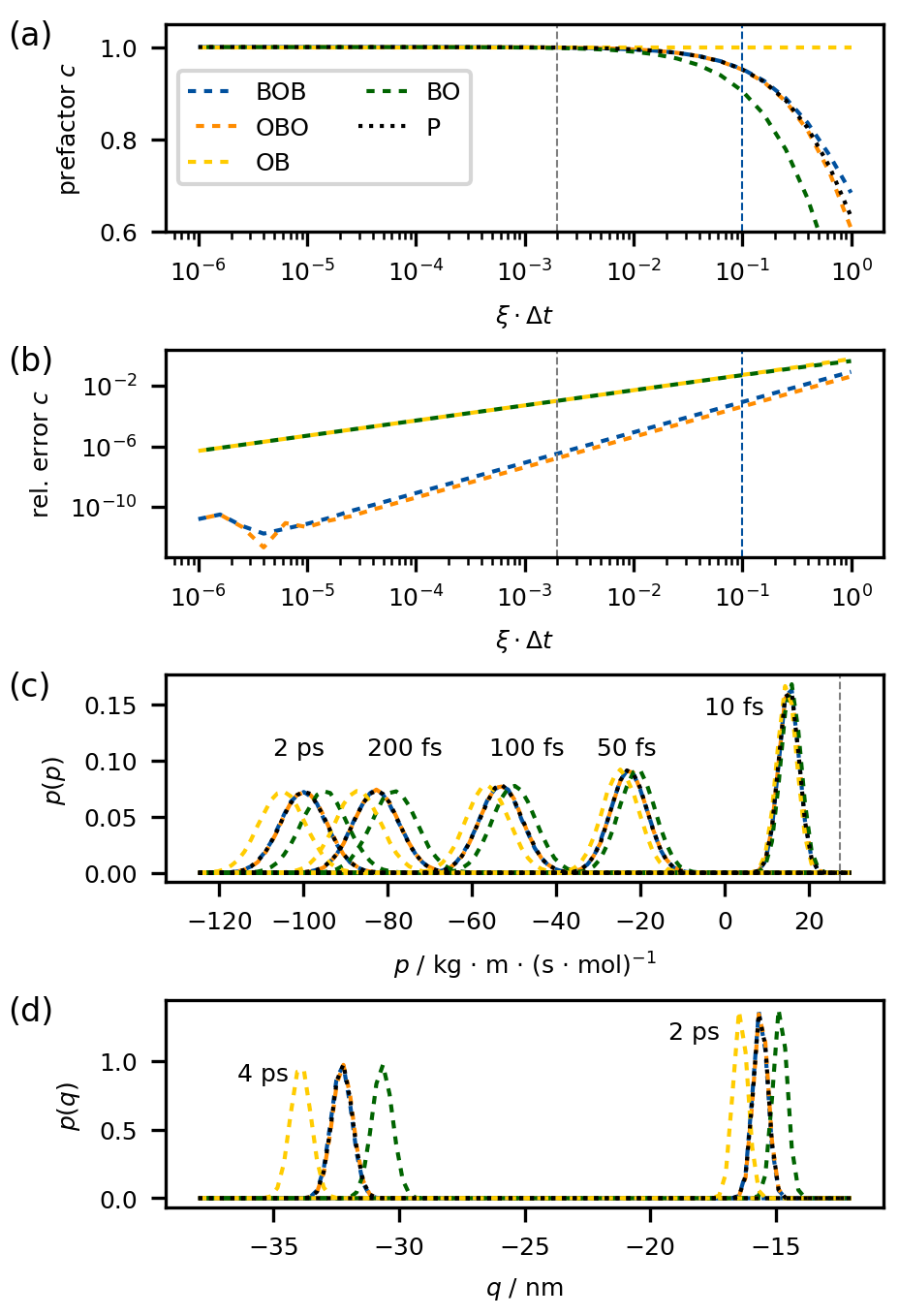}
    \caption{
    a) Prefactor $c$  (eq.~\ref{eq:generalMomentumUpdate}) that scales the force kick in eqs.~\ref{eq:BOB}-\ref{eq:BO} compared to the exact force-kick prefactor in $\mathcal{P}$ (eq.~\ref{eq:P}).  
    b) Relative error of $c$ compared to $\mathcal{P}$.
    The vertical dotted line indicates $\xi \Delta t = 0.002$, corresponding to typical MD settings of $\Delta t = 2\, \mathrm{fs}$ and $\xi = 1\, \mathrm{ps}^{-1}$.
    c) Time evolution of the momentum distribution with simulation parameters
    $m =  12\, \mathrm{g}/\mathrm{mol}$,
    $T=300\, \mathrm{K}$,
    $\xi = 10\, \mathrm{ps}^{-1}$, 
    $\Delta t = 10\, \mathrm{fs}$, 
    $\kappa = 1000\, \mathrm{kJ}/ (\mathrm{mol} \cdot \mathrm{nm})$.
    The vertical dotted line indicates the initial condition $p_0$.
    d) Time evolution of the position distribution.
    }
    \label{fig:linearPotential}
\end{figure}

Figures~\ref{fig:linearPotential}a and b show the prefactor $c$ and the error relative to $\mathcal{P}$ as a function of $\xi\Delta t$ over several orders of magnitude. 
For the BOB and OBO momentum updates, the prefactor difference, $(c_{\mathrm{scheme}}- c_P)$,  where $c_P$ is the prefactor of $\mathcal{P}$, is of second order with respect to $\xi \Delta t$ \seeSI.
Consequently, their prefactors remain very close to that of the exact update $\mathcal{P}$, even for large values of $\xi\Delta t$.  
This behavior is expected, since BOB and OBO correspond to Strang splittings \cite{Strang:1968} of the momentum update.
By contrast, the relative error of the BO and OB momentum updates is several orders of magnitude larger than that of the symmetric momentum updates.
Because the BO and OB momentum updates introduce significant errors, schemes constructed from these updates (Fig.~\ref{fig:LangevinSplitting_01}d) show correspondingly reduced accuracy compare to their symmetric counterparts (Fig.~\ref{fig:LangevinSplitting_01}b and c).
Nevertheless, for typical choices of $\Delta t$ and $\xi$ (vertical gray line in Fig.~\ref{fig:linearPotential}), the relative error of the non-symmetric momentum updates is still small - only of order $10^{-4}$.
This motivates a closer analysis of how these differences in the accuracy of the momentum updates play out in the resulting momentum distributions.
Consider a linear potential $V(q) = \kappa q$. 
Here, the force kick $b(q_k) = -\Delta t\, \kappa$ is position-independent, so the momentum update is independent of $q_k$
and can be analyzed separately from the position update.
With initial condition $p_0$, the momentum after $n$ iterations with the general momentum update in eq.~\ref{eq:generalMomentumUpdate} is normally distributed \seeSI: 
$p_n \sim \mathcal N(\mu_n,\sigma_n^2)$ with
\begin{align}
    \mu_n = d^n p_0 - c\cdot \kappa \Delta t  \frac{1-d^n}{1-d},
    \quad
    \sigma_n^2 = f^2\,\frac{1-d^{2n}}{1-d^2} .
\end{align}
Thus, all momentum update schemes discussed so far produce the same time-dependent variance $\sigma_n^2$, which converges to the variance of the Maxwell-Boltzmann distribution $\sigma_{n\rightarrow\infty}^2 = k_BTm$.
As the systems accelerate along the linear potential, their mean momentum in the direction of the force increases until it reaches a terminal value
\begin{align}
    \mu_{n\rightarrow\infty} &=  c\cdot \kappa \Delta t  \frac{1}{1-d}.
\end{align}
This is illustrated in Fig.~\ref{fig:linearPotential}.c.
The simulations use the same test particle and initial conditions as in  Fig.~\ref{fig:O_step}.
The distributions at 2 ps correspond to the terminal momentum distributions $p_n \sim \mathcal N(\mu_{n\rightarrow \infty},\sigma_{n\rightarrow \infty}^2)$. 
While the distributions obtained with the P, BOB and OBO updates are practically indistinguishable from the analytical solution, the BO distribution lags behind and the OB distributions of OB advances ahead.
Since the position updates are based in these momentum distributions, the position distributions of BO and OB increasingly deviate from the exact result with time (Fig.~\ref{fig:linearPotential}.d)
The different speeds at which the trajectory ensemble drifts down the linear potential are referred to as \emph{terminal drift} in Ref.~\citenum{Sivak:2014}.
This work suggests rescaling the time step in the A and B updates to correct for differences in the terminal drift.

At typical MD settings of $\Delta t = 2\, \mathrm{fs}$ and $\xi = 1\, \mathrm{ps}^{-1}$ (gray vertical line in Fig.~\ref{fig:linearPotential}.a and b), difference in the terminal drift across different Langevin schemes is too small to be visualized. 
Fig.~\ref{fig:linearPotential}c and d therefore uses larger settings (blue vertical line in Fig.~\ref{fig:linearPotential}.a and b) to achieve a clearly visible drift.
The force constant is representative of forces that occur in bond vibrations.

%
%
\subsection{Momentum update in a harmonic potential}
\begin{figure}
    \centering
    \includegraphics[width=\columnwidth]{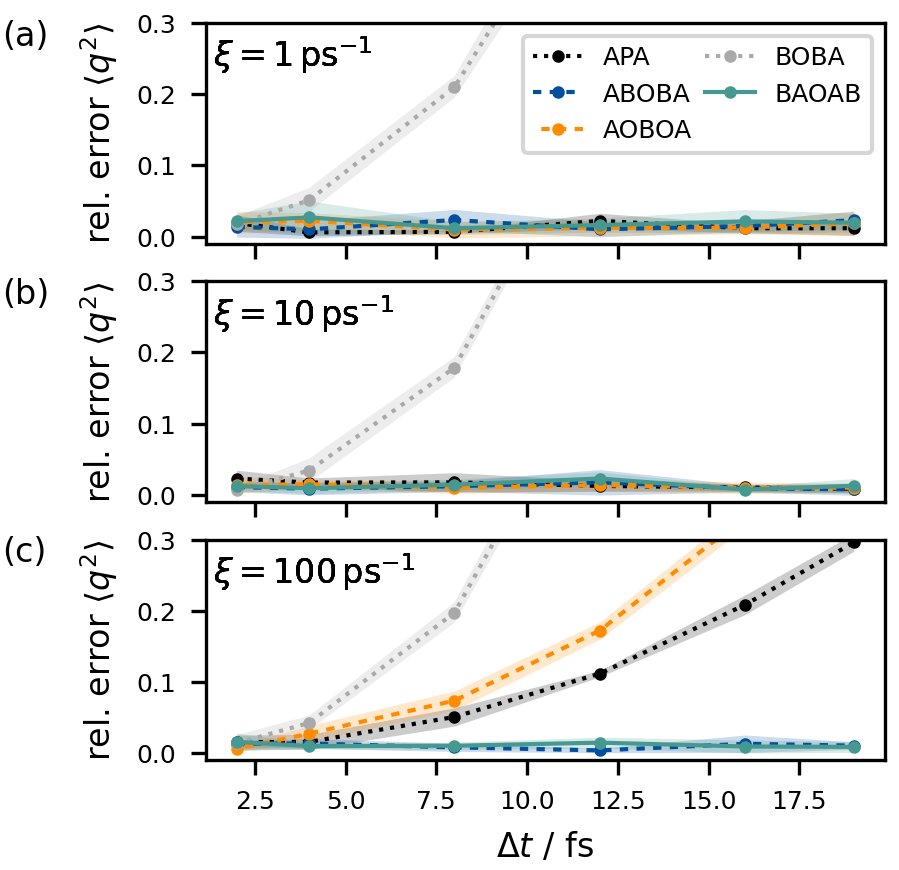}
    \caption{Relative error of $\langle q^2 \rangle$ in a harmonic potential $V(q) = \frac{1}{2}\kappa x^2$ with $\kappa = 1.2\cdot 10^{5}\mathrm{kJ}/ (\mathrm{mol} \cdot \mathrm{nm}^2)$ with simulation parameters: $m =  12\, \mathrm{g}/\mathrm{mol}$, $T=300\, \mathrm{K}$, and 
    (a): $\xi = 1\, \mathrm{ps}^{-1}$,
    (b): $\xi = 10\, \mathrm{ps}^{-1}$,
    (c): $\xi = 100\, \mathrm{ps}^{-1}$.
    }
    \label{fig:harmonicPotential}
\end{figure}

Linear potentials represent the first-order local approximation of a general potential. 
In higher-order approximations, the momentum update couples to the position update, and inaccuracies in the momentum update therefore propagate directly into the position dynamics.
First order ABO schemes, that perform poorly already for linear potentials are unlikely to be reliable in more realistic settings.
The discussion therefore now focuses on the second-order ABO schemes 
First, I consider full-step schemes which record positions at half-time steps, i.e.
$\mathcal{A'}\mathcal{P}_{\mathrm{gen}} [\mathcal{A}\mathcal{P}_{\mathrm{gen}}]^{n-1}\mathcal{A'} = [\mathcal{A}'\mathcal{P}_{\mathrm{gen}}\mathcal{A}']^n$, 
where the momentum update $\mathcal{P}_{\mathrm{gen}}$ is calculated via P, BOB or OBO.
The test system is again a single carbon atom at 300 K moving in a harmonic potential $V(q) = \kappa q^2$ 
with $\kappa = 1.2 \cdot 10^5\, \mathrm{kJ}/ (\mathrm{mol} \cdot \mathrm{nm}^2)$.
This corresponds to a relatively stiff harmonic potential, representative of a covalent C--C bond stretching mode.
The second-order approximation around a potential minimum is that a harmonic potential.
This potential is a stringent stress test for Langevin integrators because the stationary distribution induced by the discrete-time integrator can be computed analytically and compared directly to the exact continuum distribution.
The general procedure for deriving the stationary moments $\langle q^2\rangle$, $\langle p^2\rangle$, and $\langle q p\rangle$ is described in Refs.~\citenum{Leimkuhler:2013} and \citenum{leimkuhler2015molecular}. 
Explicit results for Langevin splitting schemes have been summarized in Ref.~\citenum{Sivak:2014}.
Fig.~\ref{fig:harmonicPotential} presents numerical results for $\langle q^2\rangle$.
At low and intermediate friction (Fig.~\ref{fig:harmonicPotential}a and b), the schemes of the form $[\mathcal{A}'\mathcal{P}_{\mathrm{gen}}\mathcal{A}']^n$ produce highly accurate position distributions over a wide range of time steps.
Within numerical accuracy, the error in $\langle q^2\rangle$ is essentially zero.
The time step can be increased up to the stability limit of the force calculation without a noticeable loss of accuracy.
At high friction, however, a qualitative difference between the schemes emerges.
While the error increases for the APA and AOBOA schemes, the ABOBA scheme continues to exhibit vanishing error.
Schemes of the form $[\mathcal{A}'\mathcal{P}_{\mathrm{gen}}\mathcal{A}']^n$ implicitly assume that the force remains constant during the momentum update.
For a linear potential, where this assumption is exact, the P update is exact.
For more general potentials, however, the force depends on the position and changes during the momentum update.
Close to a potential minimum, the force scales approximately linearly with the position, corresponding to a harmonic potential.
For this case, it can be shown that the BOB momentum update preserves the correct stationary configurational distribution \cite{Leimkuhler:2013, Leimkuhler:2012, Sivak:2014}.
Eq.~\ref{eq:generalMomentumUpdate} provides a second viewpoint on how the momentum update influences the position distribution. 
In a linear potential, variations in the prefactor $c$ effectively changes the slope of the potential and therefore affect the drift.
In a harmonic potential, the prefactor effectively modifies the stiffness of the potential and therefore different momentum update schemes lead to different stationary distributions in curved potentials.

\begin{figure}
    \centering
    \includegraphics[width=0.9\columnwidth]{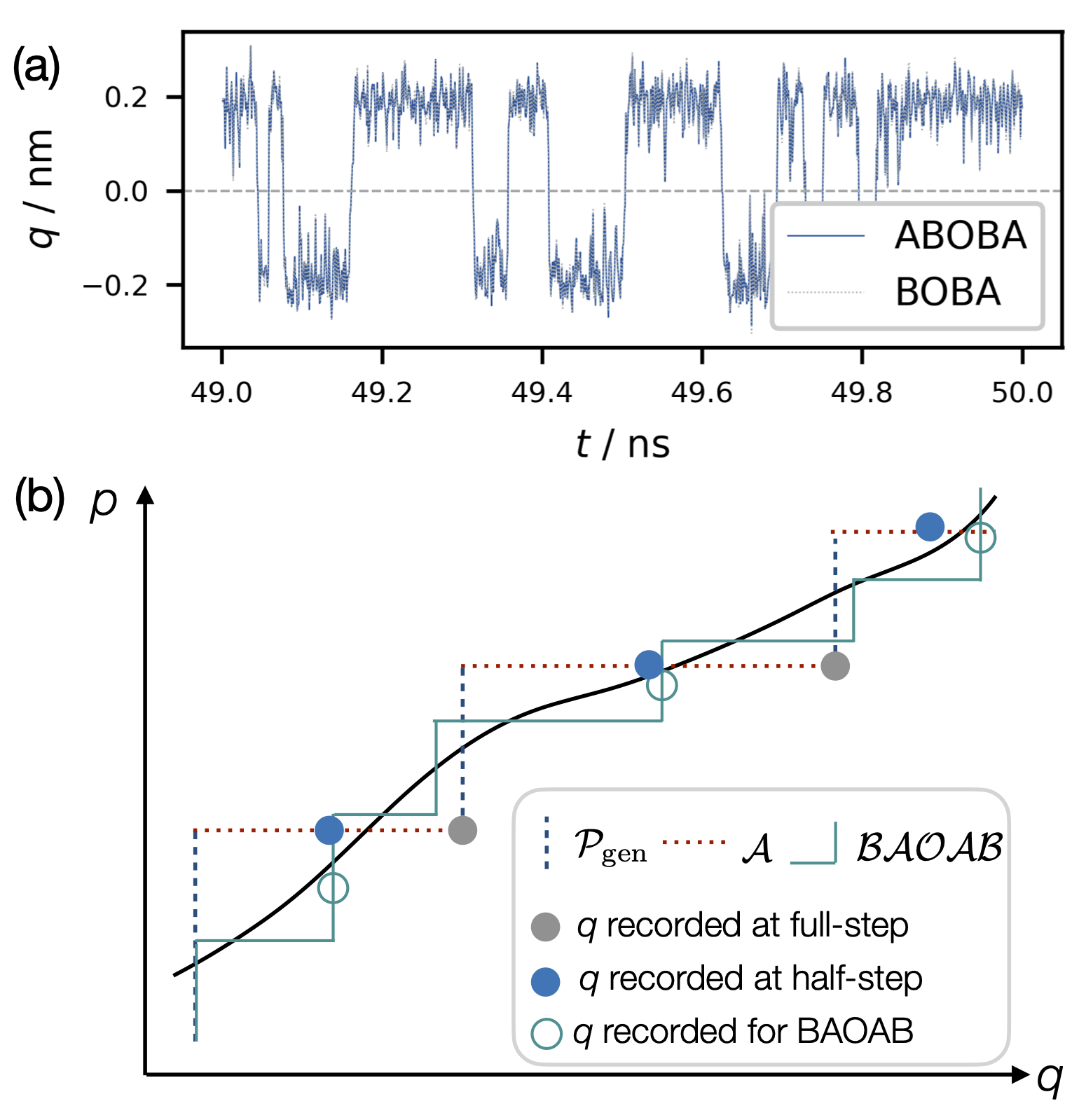}
    \caption{(a) Trajectories produced by ABOBA and BOBA. The system is the particle in the untilted double-well potential presented in Fig.~\ref{fig:doubleWell}. (b) Sketch of the path through phase space for $\mathcal{A}$-half-step and $\mathcal{A}$-half-step schemes.}
    \label{fig:ABOBA_vs_BOBA.png}
\end{figure}

\subsection{Position udpate}
\label{sec:Position_update}
So far, I focused on momentum updates; I now turn to the position update.
Since there is only a single elementary position update operator, no alternative compositions arise at this level.
However, the point within the integration cycle at which positions are recorded turns out to play an important role.
The symmetric algorithms in Fig.~\ref{fig:LangevinSplitting_01}b and c differ from their non-symmetric counterparts in a single respect: positions are written to file after half-step position updates (yellow line), whereas in the non-symmetric schemes they are recorded at full time steps (dotted red line).
Yet, changing the observation point leads to a noticeable deterioration in the accuracy of the measured position distribution.
This effect is illustrated for the BOBA scheme in Fig.~\ref{fig:harmonicPotential} (gray line).

This result is surprising, since ABOBA and BOBA generate the same internal trajectory.
Figure~\ref{fig:ABOBA_vs_BOBA.png}.a shows the final nanosecond of two 50~ns trajectories of a particle in a double-well potential, generated using the ABOBA and BOBA schemes, respectively.
The time step is $\Delta t = 40$~fs, positions are recorded every 1~ps and same random number sequence is used for both simulations.
Even with large time steps, long simulation times, and multiple barrier crossings, the two simulations remain indistinguishable. 
The difference in the accuracy of the measured distributions therefore does not originate from differences in the integration scheme itself, but from the different observation points.
Although the resulting differences between the recorded trajectories are below the visual resolution of the figure, they are systematic and thus lead to a systematic bias in the measured distributions.
This can be understood by visualizing the time-discrete path through phase space generated by the Langevin scheme (Fig.~\ref{fig:ABOBA_vs_BOBA.png}.b).
The black curve represents the reference path obtained in the limit of vanishing time step.
The ABOBA and BOBA schemes approximate this path by alternating full updates in momentum space (blue dashed segments) and full updates in position space (red dotted segments).
For sufficiently small time steps, the discrete updates remain close to the reference path, whereas increasing the step size leads to larger deviations.
Importantly, the steps of a refined (finer) discretization lie within those of a coarser discretization.
Recording positions at half steps therefore corresponds to sampling the trajectory at points that are centered within each discrete update cycle, whereas recording positions at full steps corresponds to a right-endpoint approximation of the position update.
Figure~\ref{fig:ABOBA_vs_BOBA.png}.b does not constitute a mathematical proof, but provides an intuitive geometric picture that helps rationalize the observed convergence behavior.
\subsection{$\mathcal{A}$-half-step schemes}
In the previous sections, it was established that elementary update operators can
be split and merged without changing the resulting integrator
(eq.~XX).
The following discussion addresses whether elementary update operators can also be reordered.
Reordering two update operators changes the internal structure of the algorithm unless the corresponding operators commute.
For example, applying a thermostatting step $\mathcal O$ followed by a force
update $\mathcal B$ yields a different result than applying the same steps in
reverse order, as can be seen by comparing eqs.~\ref{eq:BO} and \ref{eq:OB}.
Accordingly, the commutator $[\mathcal B,\mathcal O]=\mathcal B\mathcal O-
\mathcal O\mathcal B$ is nonzero.
The same holds for all other pairs of distinct elementary update operators \seeSI.
It is therefore impossible to rearrange the algorithms in which the position and momentum updates are advanced by effective half-steps (Fig.~\ref{fig:LangevinSplitting_01}a) such that the two position updates are consecutive and constitute an effective full-step update (Fig.~\ref{fig:LangevinSplitting_01}b and d). 
Thus, $\mathcal A$-half-step schemes are truly distinct from $\mathcal A$-full-step schemes.
In particular, the $\mathcal A$-half-step scheme BAOAB has been found to be highly accurate. 
This is confirmed in Fig.~\ref{fig:harmonicPotential}, where the numerical error in
$\langle q^2 \rangle$ is effectively zero across all investigated time steps and friction regimes.
This high accuracy of BAOAB can be understood by visualizing time-discrete path through phase space generated by the scheme (solid teal line in Fig.~\ref{fig:ABOBA_vs_BOBA.png}.b). 
Internally, BAOAB alternates half-step updates of the position with full-step updates
of the $\mathcal B$ and $\mathcal O$ operators.
Compared to the ABOBA algorithm at the same time step, this effectively corresponds
to a trajectory resolved at half the step size.
Whether the positions are written to file at the beginning (BAOA) or at the midpoint of the $\mathcal B$ update (BAOAB) does not affect the accuracy of the measured position distribution, because the B-step only updates the momenta. 
However, it is known to influence the accuracy of the measured momentum distribution.

\subsection{Double-well potential}
\begin{figure}
    \centering
    \includegraphics[width=0.95\columnwidth]{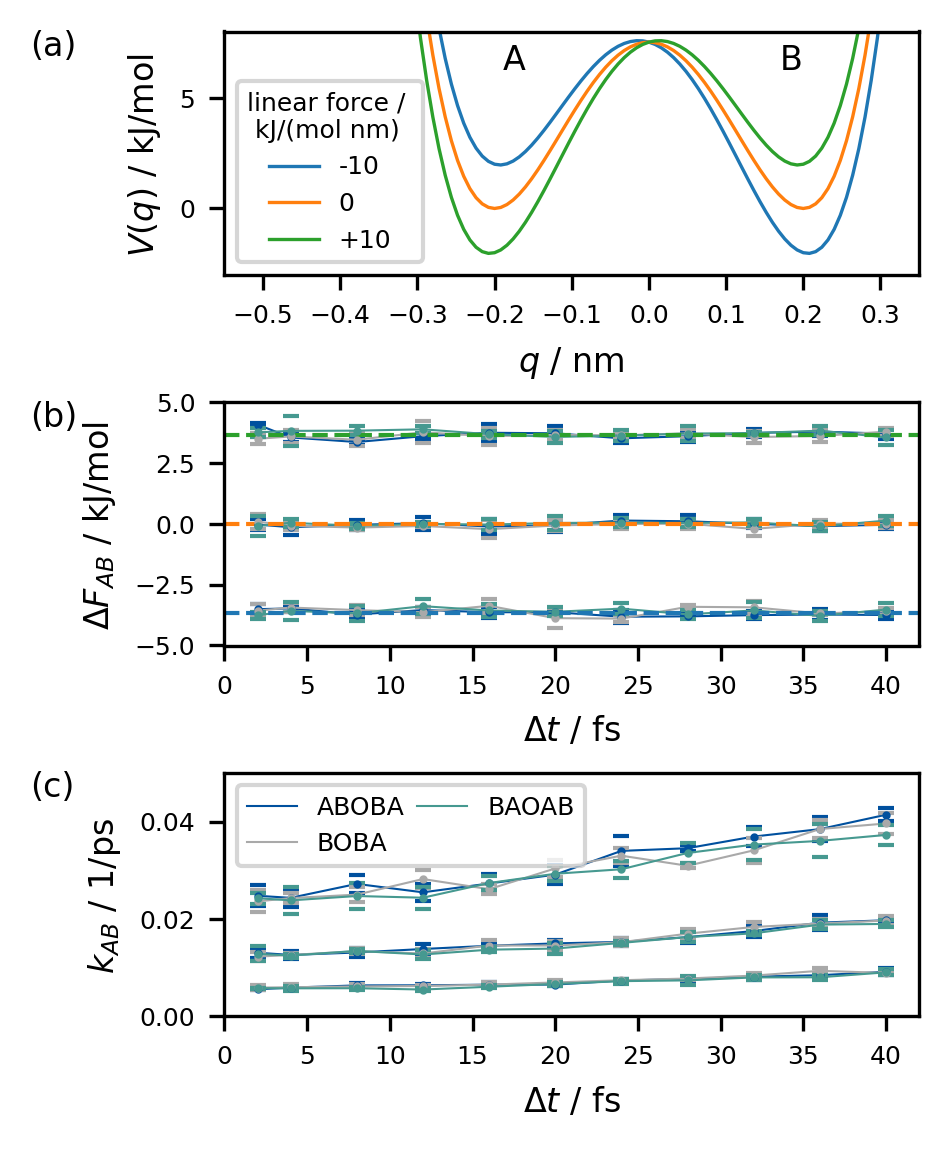}
    \caption{
    (a) Tilted double well potential. Simulation parameters are $m =  12\, \mathrm{g}/\mathrm{mol}$, $T=300\, \mathrm{K}$, and 
    $\xi = 100\, \mathrm{ps}^{-1}$.
    (b) free-energy difference between state A and B
    (c) transition rate constant A $\rightarrow$ B.}
    \label{fig:doubleWell}
\end{figure}
This naturally leads to the question of whether these accuracy differences matter in typical molecular-dynamics applications.
Figs.\ref{fig:harmonicPotential} and \ref{fig:ABOBA_vs_BOBA.png} illustrates a remarkable feature  of Langevin splitting schemes. 
Although the BOBA scheme exhibits a noticeably reduced accuracy in the measured position distribution compared to the highly accurate ABOBA scheme, both algorithms generate the same internal trajectory.
As a consequence, fluctuations within the minima and transitions across the barrier are unchanged.
In particular, both schemes spend the same amount of time on either side of the barrier, implying that free-energy estimates are largely insensitive to the choice of algorithm.
Moreover, since the configurational accuracy of ABOBA is essentially independent of the time step, this robustness is inherited by BOBA at the level of free energies, which therefore remain stable and accurate even for comparatively large time steps.

This behavior is indeed observed in the numerical experiments.
Fig.~\ref{fig:doubleWell}.a shows three double-well potentials: one without tilt (orange), one tilted to the left (green), and one tilted to the right (blue).
The minima are located at approximately $\pm 0.2\,\mathrm{nm}$; for reference, the van der Waals radius of carbon is $0.17\,\mathrm{nm}$.
For the untilted potential, the barrier height is approximately $3\,k_B T$, with the test particle again being a carbon atom at $300\,\mathrm{K}$.
The friction coefficient is set to a large value of $\xi = 100\,\mathrm{ps}^{-1}$.
Figure~\ref{fig:doubleWell}.b shows estimates of the free-energy difference between the left and right minima obtained using the ABOBA, BOBA, and BAOAB schemes for time steps ranging from $2\,\mathrm{fs}$ to $40\,\mathrm{fs}$.
Across the entire range of time steps, the estimates remain stable and in close agreement with the analytical reference value (dashed line).

Transition rates depend linearly on the population of the transition-state region, in contrast to free energies, which depend logarithmically on equilibrium probabilities.
As a result, transition rates are inherently more sensitive to small inaccuracies in the sampled position distribution \cite{Leimkuhler:2013}.
This increased sensitivity is reflected in Fig.~\ref{fig:doubleWell}.c, where the transition-rate estimates for the double-well system exhibit a slight systematic drift with increasing time step.
As before, this effect becomes noticeable only for deliberately large friction coefficients and time steps; under standard molecular-dynamics conditions, no such drift is observed.
(Note that the transition rates are estimated from direct transition counts rather
than from fluxes across the barrier, and therefore probe the accuracy of the sampled
position distribution rather than the momentum distribution.)
\section{Conclusion}
This work compared Langevin splitting schemes from the perspective of the \emph{internal trajectories} generated by the update operators.
At this level, splitting and merging identical elementary udpate operators leaves the scheme
unchanged, whereas reordering distinct steps changes the Langevin scheme. 
These observations lead to a natural classification of second-order splittings by
the operator that retains a half-step character upon iteration, yielding three
structural classes ($\mathcal A$-, $\mathcal B$-, and $\mathcal O$-half-step schemes). 
$\mathcal B$-, and $\mathcal O$-half-step schemes traverse the phase space by iterating full-step updates in the position and full-step udpates in the momentum space and can therefore be compared by analyzing the accuracy of their momentum update. 
Algorithms with equivalent operator sequence and therefore produce the same internal trajectory may differ in the observation point in the cylce at which positions are written to file - and differences in this observation point can influence the measured distributions. 
The analysis in this work focused primarily on the accuracy of the sampled positions.
An analogous analysis can be carried out for the momenta.
As a general rule of thumb, momentum distributions are closest to the Maxwell–Boltzmann distribution when momenta are observed immediately after the thermostatting step.
Overall, modern Langevin integrators are highly stable under standard molecular-dynamics conditions, and such accuracy differences typically become apparent only for deliberately large time steps and friction coefficients.


\section*{Acknowledgement}
This research has been funded by the Volkswagen Foundation through a Momentum grant. 

\section*{Supplementary material}
The supplementary material contains additional derivations and detail on computational methods.

\section*{References}
\bibliography{literature}

%
%
\newpage
\appendix

\end{document}


\setlength\parindent{0pt}


\title[]{SI: Internal Trajectories and Observation Effects in Langevin Splitting Schemes}%

\author{Bettina G. Keller}
\email[]{bettina.keller@fu-berlin.de}
\affiliation{Department of Biology, Chemistry, Pharmacy, Freie Universit\"{a}t Berlin, Arnimallee 22, D-14195 Berlin, Germany}

\date{\today}


\begin{abstract}
This is an abstract.
\end{abstract}

\maketitle

\section{Operator relations}
\label{sec:Operator_relations}

%
%
\subsection{Successive half-step updates}

\paragraph{A-step}
\begin{align}
        \mathcal{A'}\mathcal{A}' \binom{q_k}{ p_k}
    &= \mathcal{A'} \binom{q_k +a'p_k}{ p_k}
    =   \binom{q_k +a'p_k + a' p_k}{ p_k}
    =   \mathcal{A} \binom{q_k}{p_k}\, 
\end{align}
with
$a'p_k + a' p_k 
= \frac{\Delta t}{2} \frac{1}{m}p_k + \frac{\Delta t}{2} \frac{1}{m} p_k
=   \Delta t \frac{1}{m}p_k 
=  a p_k$ \, .

\paragraph{B-step}
\begin{align}
        \mathcal{B'}\mathcal{B}' \binom{q_k}{p_k}
    &=  \mathcal{B'} \binom{q_k}{p_k + b'(q_k)}
    =   \binom{q_k}{p_k + b'(q_k) + b'(q_k)}
    =   \mathcal{B}\binom{q_k}{p_k}\, 
\end{align}
with
$ b'(q_k) + b'(q_k)
= -\frac{\Delta t}{2}\nabla V(q_k)  -\frac{\Delta t}{2}\nabla V(q_k)
=   -\Delta t\nabla V(q_k) 
=  b(q_k)$.

\paragraph{O-step}
\begin{align}
    &\mathcal{O'}\mathcal{O}' \binom{q_k}{p_k}= \cr
    & \mathcal{O'} \binom{q_k}{d' \, p_k + f'\, \eta_k^{(1)}}
    =  \binom{q_k}{d'd' \, p_k + d'f'\, \eta_k^{(1)} + f'\, \eta_k^{(2)}}
    = \binom{q_k}{ d \, p_k + f'(d'\, \eta_k^{(1)} +  \eta_k^{(2)})}
\end{align}
with $d'd' = \exp(-\xi\frac{\Delta t}{2}) \exp(-\xi\frac{\Delta t}{2}) = \exp(-\xi\Delta t) = d$.

\paragraph{P-step}

\begin{align}
    \mathcal{P}'\mathcal{P}'\binom{q_k}{p_k}
    &=
    \mathcal{P}'
    \binom{q_k}{\frac{2(1 - d')}{\xi\Delta t} \, b'(q_k)+ d' p_k  + f' \, \eta_k^{(1)}} \cr
    &=
    \binom{q_k}{
        \frac{2(1 - d')}{\xi\Delta t} \, b'(q_k)+
        d'  \left(\frac{2(1 - d')}{\xi\Delta t} \, b'(q_k)+ d' p_k  + f' \, \eta_k^{(1)} \right)
        + f' \, \eta_k^{(2)} 
   } \cr
    &=
    \binom{q_k}{
        2\frac{1 - d' + d' - d'd'}{\xi\Delta t} \, b'(q_k)+ dp_k  + f' d'  \, \eta_k^{(1)}
        + f' \, \eta_k^{(2)} 
    } \cr 
    &=
     \binom{q_k}{
        \frac{1 - d}{\xi\Delta t} \, b(q_k)+ dp_k  + f' (d'  \, \eta_k^{(1)} +  \eta_k^{(2)}) 
    } \, ,
\end{align}
with $d'd' = d$ and $2b'(q_k) = b(q_k)$.

As for the $\mathcal{O}$, one can replace the two random numbers with an effective single random number $f\eta_k \stackrel{d}{=} f'd' \eta_k^{(1)} + f'\eta_k^{(2)}$. 
On the level of distributions, combining two $\mathcal{P}$-half steps is thus the same as performing an update by a single full step
\begin{eqnarray}
    \mathcal{P}'\mathcal{P}'\begin{pmatrix} q_k \\p_k\end{pmatrix}
    & \stackrel{d}{=}
     \begin{pmatrix}q_k\\ \frac{1 - d}{\xi\Delta t} \, b(q_k)+ dp_k  + f\eta_k \end{pmatrix} 
    = \mathcal{P}\begin{pmatrix} q_k \\p_k\end{pmatrix} \, .
\end{eqnarray}

%
%
\subsection{Commutators}

\paragraph{Commuting A-step and B-step}
\begin{align}
    [\mathcal{A}, \mathcal{B}] \binom{q_k}{p_k}
    &= \mathcal{A}\mathcal{B}\binom{q_k}{p_k} - \mathcal{B}\mathcal{A}\binom{q_k}{p_k}
    = \binom{q_k + ap_k+ ab(q_k)}{p_k+b(q_k)} - 
      \binom{q_k+a p_k}{p_k + b(q_k+a p_k)}  \nonumber\\[0.1cm]  
    &= \binom{ab(q_k)}{b(q_k) - b(q_k+a p_k)}
\end{align}

\paragraph{Commuting A-step and O-step}
\begin{align}
    [\mathcal{A}, \mathcal{O}] \binom{q_k}{p_k}
    &= \mathcal{A}\mathcal{O}\binom{q_k}{p_k} - \mathcal{O}\mathcal{A}\binom{q_k}{p_k}
    = \binom{q_k + a d \, p_k + a f\, \eta_k}{d \, p_k + f\, \eta_k}- 
      \binom{q_k+a p_k }{d\, p_k + f\, \eta_k} \nonumber\\[0.1cm]         
    &= \binom{ a (d-1) \, p_k + a f\, \eta_k}{0}   
\end{align}

\paragraph{Commuting O-step and B-step}
\begin{align}
    [\mathcal{B}, \mathcal{O}] \binom{q_k}{p_k}
    &=
    \mathcal{B} \mathcal{O} \binom{q_k}{p_k} - \mathcal{O} \mathcal{B} \binom{q_k}{p_k}
    =
    \binom{q_k}{dp_k + f\eta_k + b(q_k)} - \binom{ q_k}{dp_k+db(q_k) + f\eta_k} \nonumber\\[0.1cm]    
    &=
    \binom{0}{(1-d) \cdot b(q_k)} 
\label{eq:BO_commutator}    
\end{align}

\paragraph{Commuting A-step and P-step}
\begin{align}
    [\mathcal{A}, \mathcal{P}] \binom{q_k}{p_k}
    &=
    \mathcal{A}\mathcal{P}\binom{q_k}{p_k}
    -
    \mathcal{P}\mathcal{A}\binom{q_k}{p_k}
    \nonumber\\[0.1cm]
    &=
    \binom{q_k + a\!\left(\frac{1-d}{\xi\Delta t} b(q_k) + dp_k + f\eta_k\right)}
        {\frac{1-d}{\xi\Delta t} b(q_k) + dp_k + f\eta_k}
    -
    \binom{q_k + ap_k}{\frac{1-d}{\xi\Delta t} b(q_k + ap_k) + dp_k + f\eta_k}
    \nonumber\\[0.1cm]
    &=
    \binom{a\frac{1-d}{\xi\Delta t} b(q_k) + a(d-1)p_k + af\eta_k}{\frac{1-d}{\xi\Delta t}\!\left[b(q_k) - b(q_k + ap_k)\right]}
\end{align}

%
%
\subsection{Symmetrized commutators}

\paragraph{Symmetrized commutator A-step and B-step}
\begin{align}
    &\mathcal{A}' \mathcal{B}\mathcal{A}' \binom{ q_k }{p_k} -
    \mathcal{B}' \mathcal{A}\mathcal{B}' \binom{ q_k }{p_k} \nonumber\\[0.1cm]
    &=
    \binom{ q_k + 2a' p_k + a' b(q_k + a' p_k) }{ p_k + b(q_k + a' p_k) }
    -
    \binom{ q_k + a\!\left(p_k + b'(q_k)\right) }{ p_k + b'(q_k) + b'\!\left(q_k + a\!\left(p_k + b'(q_k)\right)\right) } \nonumber\\[0.1cm]
    &=
    \binom{ 2a'p_k + a' b(q_k + a' p_k) - a\!\left(p_k + b'(q_k)\right)}
          { b(q_k + a' p_k) - b'(q_k) - b'\!\left(q_k + a\!\left(p_k + b'(q_k)\right)\right) } \, .
\end{align}

\paragraph{Symmetrized commutator A-step and O-step.}
\begin{align}
    &\mathcal{A}' \mathcal{O}\mathcal{A}' \binom{ q_k }{p_k} - \mathcal{O}' \mathcal{A}\mathcal{O}' \binom{ q_k }{p_k} \cr
    &= \binom{ q_k + a'(1+d)\,p_k + a'f\,\eta_k }{ d\,p_k + f\,\eta_k } - \binom{q_k + ad'p_k + a f'\eta_k^{(1)} }{dp_k +f'(d'\eta_k^{(1)} + \eta_k^{(2)}) } \cr
    &= \binom{ (a'+a'd - ad')\,p_k + a'f\,\eta_k -a f'\eta_k^{(1)}}{  f\,\eta_k - f'(d'\eta_k^{(1)} + \eta_k^{(2)})} \cr
    &\stackrel{d}{=} \binom{ (a'+a'd - ad')\,p_k + a'f\,\eta_k -a f'\eta_k^{(1)}}{ 0} 
\end{align}
where we replaced $f\eta_k \stackrel{d}{=} f'd'\eta_k^{(1)} + f'\eta_k^{(2)}$ in the last line.

\paragraph{Symmetrized commutator O-step and B-step}
\begin{align}
    \mathcal{B}' \mathcal{O}\mathcal{B}' \binom{ q_k }{p_k} - \mathcal{O}' \mathcal{B}\mathcal{O}' \binom{ q_k }{p_k} 
    &\stackrel{d}{=} \binom{q_k }{\frac{d+1}{2} b(q_k) + dp_k + f \eta_k} - \binom{q_k}{ d'b(q_k) + d p_k + f \eta_k } \cr
    &= \binom{0 }{(\frac{d+1}{2} -d') b(q_k)} 
\label{eq:BO_symmetrized_commutator}        
\end{align}
where we replaced $f\eta_k \stackrel{d}{=} f'd'\eta_k^{(1)} + f'\eta_k^{(2)}$ in the last line.

\newpage
\section{Momentum update}
\label{sec:Momentum_distribution}

%
%
\subsection{Composite momentum updates}

The composite momentum updates of $\mathcal{B}'\mathcal{O}\mathcal{B}' \binom{q_k}{p_k}$ and $\mathcal{O}'\mathcal{B}\mathcal{O}' \binom{q_k}{p_k}$ are recorded in eq.~\ref{eq:BO_symmetrized_commutator}.
The composite momentum updates of $\mathcal{B}\mathcal{O} \binom{q_k}{p_k}$ and $\mathcal{O}\mathcal{B} \binom{q_k}{p_k}$ are recorded in eq.~\ref{eq:BO_commutator}.

%
%
\subsection{Commutators of momentum update operators}
The general form of the momentum update operator is
\begin{align}
    \mathcal{P}_{\mathrm{gen}} \binom{q_k}{p_k} &= \binom{ q_k}{ cb(q_k) + dp_k + f\eta_k}\, ,
\end{align}
where the prefactor $c$ depends on the specific operator.
Consider two such momentum updates $\mathcal{P}_1$ and $\mathcal{P}_2$ with prefactors $c_1$ and $c_2$.
Their composition difference is given by the commutator
\begin{align}
    &[\mathcal{P}_1, \mathcal{P}_2]\binom{q_k}{p_k}\cr
    &=  \mathcal{P}_1 \mathcal{P}_2\binom{q_k}{p_k} - \mathcal{P}_2 \mathcal{P}_1\binom{q_k}{p_k} \cr
    &=  \mathcal{P}_1 \binom{ q_k}{ c_1b(q_k) + dp_k + f\eta_k^{(1)}} - 
        \mathcal{P}_2 \binom{ q_k}{ c_2b(q_k) + dp_k + f\eta_k^{(1)}} \cr
    &=  \binom{ q_k}{ c_2b(q_k) + dc_1b(q_k) + ddp_k + df\eta_k^{(1)} + f\eta_k^{(2)}} -   
        \binom{ q_k}{ c_1b(q_k) + dc_2b(q_k) + ddp_k + df\eta_k^{(1)} + f\eta_k^{(2)}} \cr
    &=  \binom{ 0}{ (c_2-c_1)(1-d)b(q_k) }   
\end{align}
asuming that the same noise realization $\eta_k^{(1)}$ and $\eta_k^{(2)}$ is used in both operator compositions.
The accuracy difference scales with the prefactor difference $c_2-c_1$, and the order of this difference can be assessed by expanding the prefactors in powers of $x=\xi\Delta t$:
\begin{align}
    c_P &= \frac{1-d}{\xi\Delta t}
    = \frac{1-e^{-x}}{x}
    = 1 - \frac{x}{2} + \frac{x^2}{6} - \frac{x^3}{24} + \mathcal{O}(x^4), \cr
    c_{BOB} &= \frac{d+1}{2}
    = \frac{1+e^{-x}}{2}
    = 1 - \frac{x}{2} + \frac{x^2}{4} - \frac{x^3}{12} + \mathcal{O}(x^4), \cr
    c_{OBO} &= d'
    = e^{-x/2}
    = 1 - \frac{x}{2} + \frac{x^2}{8} - \frac{x^3}{48} + \mathcal{O}(x^4), \cr
    c_{OB} &= 1, \cr
    c_{BO} &= d
    = e^{-x}
    = 1 - x + \frac{x^2}{2} - \frac{x^3}{6} + \mathcal{O}(x^4).
\end{align}
Taking the exact momentum update $\mathcal{P}$ as the reference, one obtains the following leading-order differences:
\begin{align}
    c_{BOB} - c_P &= \mathcal{O}(x^2) \cr
    c_{OBO} - c_P &= \mathcal{O}(x^2) \cr
    c_{OB}  - c_P &= \mathcal{O}(x) \cr
    c_{BO}  - c_P &= \mathcal{O}(x).
\end{align}

\subsection{Constant potential}
For a constant potential, $b(q_k)=0$, and the momentum update reduces to the
Ornstein--Uhlenbeck update $\mathcal O$.
In this case, although the positions may still evolve under the integration scheme, the momentum update is independent of the position and can therefore be analyzed in isolation.
Starting from the initial condition $(q_0,p_0)$, the momentum is given by a series of update steps with the $\mathcal{O}$-operator.
For two time steps this yields
\begin{align}
    \binom{q_2}{p_2}
    &= \mathcal O \mathcal O \binom{q_0}{p_0}
    = \mathcal O \binom{q_0}{d\,p_0 + f\,\eta_1 } 
    = \binom{q_0}{d^2 p_0 + d f\,\eta_1 + f\,\eta_2}
\end{align}
The general expression for the momentum after
$n$ steps follows by iteration:
\begin{align}
    p_n
    &= d^n p_0 + \sum_{k=1}^{n} d^{\,n-k}\, f\,\eta_k .
\end{align}
The momentum $p_n$ is thus a linear combination of independent normal random variables and is therefore itself normally distributed,
\begin{align}
    p_n \sim \mathcal N(\mu_n,\sigma_n^2),
    \qquad
    \mu_n = d^n p_0,
    \qquad
    \sigma_n^2 = \sum_{k=1}^{n} \bigl(d^{\,n-k} f\bigr)^2 .
\end{align}

The variance is a geometric series and can be evaluated as
\begin{align}
    \sigma_n^2
    &= f^2 \sum_{k=1}^{n} d^{2(n-k)}
     = f^2 \sum_{l=0}^{n-1} (d^2)^l
     = f^2\,\frac{1-d^{2n}}{1-d^2},
\end{align}
where we substituted $l=n-k$.
For $n\rightarrow \infty$, the momentum distribution converges to the Maxwell-Boltzmann distribution
\begin{align}
    \mu_{n\rightarrow\infty} = 0 , \qquad
    \sigma_{n\rightarrow\infty}^2 =f^2\,\frac{1}{1-d^2} = \frac{k_BTm(1-e^{-2\xi\Delta t})}{1-e^{-2\xi\Delta t}} = k_BTm \, .
\end{align}

\subsection{Linear potential}
For a linear potential, the force does not depend on the position and therefore the momenta evolve independent of the position according to
\begin{align}     
    \dot{p} &= -\kappa - \xi p + \sqrt{2\xi k_BTm} \, \eta(t) \, ,    
\end{align}
where $\kappa$ is a force constant.
The  analytical momentum distribution for the Ornstein-Uhlenbeck process with linear potential is
\begin{align}
    p(t) \sim \mathcal N(\mu(t),\sigma^2(t)),
    \quad
    \mu(t) =  \mathrm{e}^{-\xi t}\, p_0 - \frac{\kappa}{\xi}\left(1-\mathrm{e}^{-\xi t}\right),
    \quad
    \sigma^2(t) = m k_B T \left(1-\mathrm{e}^{-2\xi t}\right).
\end{align}
Analogously, in a numerical simulation of this system, the  force kick $b(q_k) = b = -\kappa \Delta t$, where $\Delta t$ is the timestep, does not depend on the position $q_k$.
Therefore, the momentum update $\mathcal{P}_{\mathrm{gen}}$ is independent of the position and can be analyzed independent of the position update $\mathcal{A}$. 
The general form of the momentum update operator at a linear potential is
\begin{align}
    \mathcal{P}_{\mathrm{gen}} \binom{q_k}{p_k}  &= \binom{q_k}{ cb + dp_k + f\eta_k} \, .
\end{align}
Applying the momentum update twice yields
\begin{align}
    \binom{q_2}{p_2}  
    &= \mathcal{P}_{\mathrm{gen}} \mathcal{P}_{\mathrm{gen}} \binom{q_0}{p_0}  
    = \mathcal{P}_{\mathrm{gen}} \binom{q_0}{cb + dp_0 + f\eta_1}  
    = \binom{q_0}{cb + dcb + ddp_0 + df\eta_1 + f\eta_2}  \, .
\end{align}
The general expression for the momentum after $n$ steps follows by iteration:
\begin{align}
    p_n
    &= d^n p_0 + cb\sum_{k=1}^{n}  d^{n-k}  + \sum_{k=1}^{n} d^{\,n-k}\, f\,\eta_k  
    = d^n p_0 + cb \frac{1-d^n}{1-d}  + \sum_{k=1}^{n} d^{\,n-k}\, f\,\eta_k \, ,
\end{align}
where we used that the second term is a geometric series
$\sum_{k=1}^{n}  d^{n-k} = \sum_{l=0}^{n-1} d^l = \frac{1-d^n}{1-d}$ with the substitution $n-k = l$ and inverting the order of the sum.
Using an analogous argumentat as for the constant potential, $p_n$ is normally distributed with 
\begin{align}
    p_n \sim \mathcal N(\mu_n,\sigma_n^2),
    \qquad
    \mu_n = d^n p_0 + cb \frac{1-d^n}{1-d},
    \qquad
    \sigma_n^2 = f^2\,\frac{1-d^{2n}}{1-d^2} .
\end{align}
For $n\rightarrow \infty$, mean and variance converge to 
\begin{align}
    \mu_{n\rightarrow\infty} = cb \frac{1}{1-d},
    \qquad
    \sigma_{n\rightarrow\infty}^2 = k_BTm \, .
\end{align}

\newpage
\section{Computational methods}

\subsection{Figure 2 in the main part}
A particle with mass $m = 12 \, \mathrm{g}/\mathrm{mol}$ is simulated according to the Ornstein-Uhlenbeck process at temperature $T=300\, \mathrm{K}$ and friction $\xi = 1 \, \mathrm{ps}^{-1}$.
In the first numerical experiment, a swarm of $10^{5}$ trajectories is initialized at  $(q_0 = 0\, \mathrm{nm}, p_0 = 78.97\, \mathrm{kg}\, \mathrm{m} / (\mathrm{mol}\, \mathrm{s}) )$, where the initial momentum corresponds to five times the thermal momentum.
The trajectories are propagated by $\mathcal{O}^n$ with $\Delta t = 2\, \mathrm{fs}$ for $n= 5, 25, 50, 100$ iterations. 
The momenta at the simulation end points are recorded. 
New random numbers are drawn at each integration step and for each trajectory. 
The blue dashed lines in Fig.~2 show the histograms of the recorded momenta.
The experiment is repeated, but this time the trajectories are propagated by $(\mathcal{O}'\mathcal{O}')^n$, i.e.~$\mathcal{O}'$ has an effective time step of $\Delta t = 1\, \mathrm{fs}$, for  $n= 5, 25, 50, 100$ iterations.
The momenta at the simulation end points are recorded. 
New random numbers are drawn at each integration step and for each trajectory. 
The orange dotted lines in Fig.~2 show the histograms of the recorded momenta.

The analytical solution for the time evolution of the momentum distribution (gray line) is a normal distribution with time-dependent mean and standard deviation: 
\begin{align}
    p(p,t) &= \mathcal{N}\left(\exp(-\xi t), \sqrt{k_BTm(1-e^{-2\xi t})} \right)
\end{align}

\subsection{Figure 3 in the main part}

Langevin dynamics simulations were performed for a one-dimensional particle in a linear potential using an in-house Python framework. The system consisted of a particle with mass \( m = 12 \,\mathrm{g\,mol^{-1}} \) at temperature \( T = 300 \,\mathrm{K} \), propagated with collision frequency \( \xi = 10 \,\mathrm{ps^{-1}} \) and time step \( \Delta t = 0.010 \,\mathrm{ps} \). The linear potential was defined as \( V(x) = k(x-a) \) with force constant \( k = 1000 \,\mathrm{kJ\,mol^{-1}\,nm^{-1}} \) and shift \( a = 0 \,\mathrm{nm} \). Initial velocities were chosen deterministically as five times the thermal velocity to probe mean drift behavior.  

Stochastic trajectories were generated using operator-splitting Langevin integrators composed of position (\(\mathcal A\)), force (\(\mathcal B\)), Ornstein--Uhlenbeck thermostat (\(\mathcal O\)), and combined deterministic propagation (\(\mathcal P\)) updates. Schemes ABOBA, AOBOA, ABOA, AOBA, and APA) were evaluated. For each scheme and total propagation time, ensembles of \(10^5\) independent trajectories were generated. Total propagation times were varied by performing \( n = \{1, 5, 10, 20, 200, 400\} \) integration steps.  
For each trajectory, final positions and velocities were recorded and stored as NumPy arrays for subsequent statistical analysis. 

\subsection{Figure 4 in the main part}

Langevin dynamics simulations were performed for a one-dimensional particle in a harmonic potential using an in-house Python simulation framework. The particle mass was \( m = 12 \,\mathrm{g\,mol^{-1}} \) and simulations were carried out at temperature \( T = 300 \,\mathrm{K} \) with collision frequency \( \xi = \{1, 10, 100\} \,\mathrm{ps^{-1}} \). The equations of motion were integrated using a time step \( \Delta t = \{2, 4, 8, 12, 16, 18, 19\} \,\mathrm{fs} \).

The potential energy surface was defined as a quadratic potential
\begin{equation}
V(x) = \frac{1}{2} k (x-a)^2 ,
\end{equation}
with force constant \( k = 1.2 \times 10^{5} \,\mathrm{kJ\,mol^{-1}\,nm^{-2}} \) and shift \( a = 0 \,\mathrm{nm} \).

Initial positions were set to \( x_0 = 0 \,\mathrm{nm} \). Initial velocities were sampled from the one-dimensional Maxwell--Boltzmann distribution,
\begin{equation}
v_0 \sim \mathcal{N}\!\left(0, \frac{RT}{M}\right),
\end{equation}
and converted to simulation units (nm/ps).

Trajectories were propagated using the APA, ABOBA, AOBOA, BOBA and BAOAB schemes.
For each simulation condition, \( n_{\mathrm{rep}} = 5 \) independent replicas were generated.

Each replica was propagated for a total simulation time of \( T_{\mathrm{sim}} = 10\,\mathrm{ns} \). Trajectory data (positions and velocities) were stored an output interval of \(\sim 1 \,\mathrm{ps} \). The resulting trajectories were stored as NumPy arrays for subsequent statistical analysis.

\subsection{Figure 6 in the main part}
Langevin dynamics simulations were performed for a one-dimensional particle evolving in a tilted double-well potential using an in-house Python simulation framework. The particle mass was \( m = 12 \,\mathrm{g\,mol^{-1}} \) and simulations were performed at temperature \( T = 300 \,\mathrm{K} \) with collision frequency \( \xi = 100 \,\mathrm{ps^{-1}} \). The integration time step was varied \( \Delta t = \{2, 4, 8, 12 16, 20, 24, 28, 32, 36, 40   \} \,\mathrm{ps} \).

The potential energy surface was constructed as a double-well potential with an additional linear perturbation,
\begin{equation}
V(x) = k_{\mathrm{dw}}\big[(x-a)^2 - b\big]^2 + k_{\mathrm{lin}} x ,
\end{equation}
with parameters \( k_{\mathrm{dw}} = 4700 \,\mathrm{kJ\,mol^{-1}\,nm^{-4}} \), \( a = 0 \,\mathrm{nm} \), and \( b = 0.04 \,\mathrm{nm^2} \). The linear perturbation coefficient was varied as \( k_{\mathrm{lin}} \in \{-10, 0, 10\} \,\mathrm{kJ\,mol^{-1}\,nm^{-1}} \).

Initial positions were set to \( x_0 = 0 \,\mathrm{nm} \), while initial velocities were sampled from the one-dimensional Maxwell--Boltzmann distribution,
\begin{equation}
v_0 \sim \mathcal{N}\!\left(0, \frac{RT}{M}\right),
\end{equation}
and converted to simulation units (nm/ps).

Trajectories were propagated using the the schemes ABOBA, BAOAB and BOBA. For each value of the linear perturbation parameter, \( n_{\mathrm{rep}} = 5 \) independent replicas were simulated.

Each replica was propagated for a total simulation time of \( T_{\mathrm{sim}} = 50\,\mathrm{ns} \). Trajectory data (positions and velocities) were stored every $\sim$ 1 ps. The resulting trajectories were stored as NumPy arrays for subsequent statistical and kinetic analysis.

\clearpage
\bibliography{literature}